# Flexible Hardware-Enabled Guarantees for AI Compute

James Petrie, Onni Aarne, Nora Ammann, David 'davidad' Dalrymple

*April 2025*

Flexible Hardware-Enabled Guarantees

# Part I: Overview of Flexible Hardware-Enabled Guarantees

James Petrie, Onni Aarne, Nora Ammann, David 'davidad' Dalrymple
*April 2025*

## About This Report

This is the first part of a three-part exploration of flexible hardware-enabled guarantees (flexHEGs), commissioned by ARIA.[1] This first part provides a comprehensive introduction of the flexHEG security system, including an overview of the governance and security capabilities it offers, its potential development and adoption paths, and the remaining challenges and limitations it faces. Part II discusses in detail the technical options for implementing flexHEGs on frontier AI chips, and its challenges. This part would be of the greatest interest for readers interested in funding or directly contributing to the development of the technology. Part III discusses the governance applications of the technology at more depth, and would be of the greatest interest to policymakers and policy researchers interested in understanding how flexHEGs might be used to solve problems that they are working on.

### Acknowledgements

We would like to thank Aaron Scher, Amanda El-Dakhakhni, Andrew Critch, Asher Brass, Ben Harack, Connor Hunter, Eddie Jean, Erich Grunewald, Evan Miyazono, Gabriel Kulp, Halfdan Holm,  Jacob Lagerros, Joe O'Brien, Jonathan Happel, Joshua Clymer, Mehmet Sencan, Oliver Guest, Péter Drótos, Rose Hadshar, Valtteri Lipiäinen, and Will Hodgkins for useful discussions and comments.

---

[1] The report is based on an original concept developed by David 'davidad' Dalrymple. We expand the definition of flexHEGs somewhat compared to earlier work, including our interim report [1]. See Appendix A for an explanation of how and why, but this should not be relevant to readers who have not engaged with that prior work.

# Executive Summary

As general purpose AI technology develops, it may come to have far-reaching implications for international security, comparable to e.g. nuclear technology, creating coordination problems for states. We propose Flexible Hardware-Enabled Guarantee (flexHEG) mechanisms: security systems added to AI chips that combine a tamper-proof enclosure with a guarantee processor to enable trustworthy, privacy-preserving verification and enforcement of claims about how the chips are used. These mechanisms could be used to flexibly solve these coordination problems as they arise, by allowing states and other actors to make credible, verifiable commitments. To ensure that all parties could trust these devices, they would have to be made fully open source and auditable, as well as robust to tampering even from state-level adversaries.

Users could use the guarantee processor's capabilities to verify whatever claims about the device they want to verify, or are asked by authorities to verify. Optionally, the guarantee processor's firmware could be configured to enforce a ruleset regarding the behavior of the AI chip. Different sets of actors could be given the power to set and update this ruleset, from simply the owner of the chip up to a quorum of states.

FlexHEG devices could be used in a range of ways to address a range of different international security problems that AI technology may create in the future. Their flexibility and updateability would allow them to adapt very rapidly.

FlexHEGs could enable many concrete governance mechanisms, including:
- Verifying or restricting the approximate location of chips.
- Verifiable, mutually privacy-preserving evaluations of the capabilities and other properties of AI models.
- Securing access to model weights by requiring them to be encrypted before they can be moved off the flexHEG device, possibly in such a way that they can only be decrypted by specific other flexHEG devices. This would allow controlled deployment, e.g. only allowing a model to be deployed with particular safeguards in place.
- Verification or limits on total training compute or other properties of trained AI models.
- Requiring the possession of a non-expired license to run computations with specific technical properties, e.g. exceeding some compute threshold.
- More complex rules, such as requiring a standardized evaluation protocols to be incorporated into the training process for sufficiently large training runs

Several technical problems would need to be solved to create sufficiently effective and secure flexHEGs. We are confident that some variant of this proposal is technically feasible. However, early versions may need to compromise on security and the sophistication of supported rules. In order to ensure that the technology is available by the time it is needed, concerted R&D efforts to develop production-ready flexHEGs would ideally begin as soon as possible.

# Table of Contents

# Introduction

> *"Unsafe development, deployment, or use of AI systems may pose catastrophic or even existential risks to humanity within our lifetimes. These risks from misuse and loss of control could increase greatly as digital intelligence approaches or even surpasses human intelligence.*
>
> *In the depths of the Cold War, international scientific and governmental coordination helped avert thermonuclear catastrophe. Humanity again needs to coordinate to avert a catastrophe that could arise from unprecedented technology."*
>
> — International Dialogues on AI Safety, joint statement, Beijing, March 2024

Artificial intelligence systems are becoming increasingly powerful, with many experts suggesting the technology could be among the most consequential in human history, potentially rivaling nuclear technology in its implications for international security [2], [3], [4]. This has led to governance efforts around the world [5], , and growing warnings that some form of international coordination may be needed to preserve security and stability [6].

However, major states find themselves in an intense technological competition regarding AI capabilities, making them reluctant to impose significant regulations without credible assurances that their competitors will do the same. While these risks could potentially be addressed through mutual verification of AI development practices, current approaches would require extensive transparency that could compromise national security interests and potentially exacerbate race dynamics by accelerating technology transfer between competing states [7], [8].

Ideally, this verification challenge could be solved by having a trusted third party conduct inspections and only share high-level findings about compliance, without revealing sensitive technical details. Such a third party could even enforce compliance with agreed-upon rules when necessary. However, in the current international system, finding or creating entities that all major powers would trust with such authority is extremely difficult.

This report proposes a technical solution: Flexible Hardware-Enabled Guarantee (flexHEG) mechanisms that could be added to AI chips to act as a trusted third party, enabling privacy-preserving verification and enforcement mechanisms between states. To ensure that all parties could trust these mechanisms, the flexHEG design would be fully open source and auditable. To prevent tampering, the devices would need to be housed in secure enclosures that could resist even highly-motivated nation states. The mechanisms would also be programmable and updateable, allowing governance to adapt as technology develops and new risks emerge, thus avoiding the accumulation of outdated or ineffective regulations.

FlexHEGs would enable privacy-preserving verification of a wide range of governance mechanisms, from simple non-proliferation measures to sophisticated international agreements. In addition to verification, flexHEGs could optionally be configured to automatically enforce commitments or rules. This enforcement capability would allow flexHEGs to address situations where conventional deterrence fails – whether dealing with actors like terrorist organizations that are indifferent to punishment, or powerful states that can resist it.

## Conceptual Overview of the FlexHEG Design Stack

In the flexHEG design, each of the AI chips (such as GPUs) would be placed in **a tamper-proof secure enclosure** along with an auxiliary **guarantee processor**. All of the chip's memory and other components would also be contained within the enclosure. Alternatively, multiple AI chips could be included in a single enclosure, managed by a single guarantee processor.

The guarantee processor would ideally be an open-source, standardized, general-purpose processor that sits between the AI chip and the rest of the world, able to locally access all information and instructions going to and from the chip as well as some aspects of the AI chip's state.

Crucially, the guarantee processor would be able to encrypt and authenticate data coming from the AI chip and from the guarantee processor itself, and thus enable privacy-preserving, sophisticated, and programmable verification schemes. Encryption would allow improved security through preventing anyone other than the intended recipient — such as the chip owner, a regulator, or a guarantee processor in another flexHEG device— from accessing or modifying the data sent from the guarantee processor. This would be similar to existing confidential computing systems [9]. Authenticated communication between flexHEG devices would enable verification schemes to be coordinated and scaled across arbitrary numbers of flexHEG devices, thus allowing verifiable claims about complex, distributed workloads (see further discussion in Appendix B and Part II).

FlexHEGs would enable privacy-preserving verification by allowing detailed verified claims to be aggregated across devices to produce more abstract claims such as "this model was trained with less than *X* total FLOP", without revealing exact FLOP values or other technical details.

The tamper-proof enclosure would protect both the main chip and the guarantee processor from snooping or interference. If any attempt is made to tamper with the enclosure, a tamper-detection system would be triggered, activating mechanisms that will wipe any secret information on both chips, including the private key that the chip would use to sign verifiable claims.[2] Optionally the tamper response could be configured to also render the AI chip permanently inoperable by, e.g., blowing a large number of microscopic fuses on the AI chip, thus preventing the potentially-compromised chip from being misused.

Users could use the guarantee processor's capabilities to verify whatever claims about the device they want to verify, or are asked by authorities to verify. Optionally, the guarantee processor's firmware could be configured to enforce a ruleset regarding the behavior of the AI chip, e.g. require that any models trained above a particular FLOP count also implement particular safety features. Different sets of actors could be given the power to set and update this ruleset, from simply the owner of the chip up to a quorum of states.

---

[2] The tamper-detection and response mechanisms would be made fully self-contained and separate from the guarantee processor, to ensure there is no way to remotely trigger them.

In order to keep all devices up-to-date on security updates and possible ruleset changes, guarantee processors could be required to regularly — e.g., every three months — install a firmware update, or the guarantee processor will block the chip from operating. This update interval itself would be programmable so that it can be "tightened" or "loosened" depending on what a situation calls for.

Finally, the supply chain for the enclosures and guarantee processors would need to be secured and monitored to the satisfaction of all relevant parties to ensure that they can trust the integrity of the guarantee processors.

# How FlexHEGs Could Address Risks to International Security

Several potential risks to international security can be seen as stemming from, or being exacerbated by, the international competitive dynamics discussed in the introduction. This section will outline some examples of governance mechanisms that FlexHEGs could enable to help address these risks. For more details on these mechanisms and many others, see Appendix B, and for more in-depth discussion of each of these risks, see Part III.

All of these risks are presently various degrees of speculative: We present them as examples to illustrate the potential of flexHEGs. The flexibility of the technology would allow actual mechanisms to be tailored to whatever risks prove most realistic, and adjusted as we learn more.

## Malicious Use

Future AI technology may be powerful enough to significantly empower rogue, malicious actors such as terrorists, improving their ability to develop and access e.g. dangerous cyber or biological capabilities [10].

FlexHEGs could enable several different mechanisms that would help address these risks, including:
- **Location verification:** FlexHEGs could enable automated verification of approximate AI chip locations, making it more difficult for malicious actors to divert chips to their facilities undetected [11]. Location restrictions could even be enforced, preventing chips from operating if they cannot verify their proximity to trusted landmarks.
- **Controlled deployment:** FlexHEGs could enable models to be authorized for deployment only on particular flexHEG devices, with guarantees that the model cannot be extracted and can only be used with intended misuse safeguards in place. This would allow models to be protected from malicious use and uncontrolled proliferation while enabling widespread deployment to less-trusted jurisdictions and end users, while giving those end users guaranteed, private access.

## Loss of Control

Risks from loss of control are potentially among the most severe risks from advanced AI [2], [10]. In the race for economic and military advantage, states and companies may attempt to make their AI systems as autonomous as possible, reducing the need for human oversight. Without sufficient safeguards, such systems could escape human control entirely, with catastrophic or even existential consequences.

FlexHEGs could alleviate this dangerous race dynamic by allowing states to provide credible assurances that they are developing and deploying their systems responsibly.

At some point in the future, AI capabilities may reach a point where the acute risk of loss of control appears to be high if progress to make AI systems more capable and autonomous continues. In such a situation, states may be interested in providing temporary guarantees that they will not create new models exceeding existing capabilities. One simple way to operationalize such a guarantee would be in terms of **training compute limits**: flexHEGs would allow states to make verifiable claims about how much total computing capacity was used to train a particular model, and even that no model exceeding some compute threshold has ever been trained on a particular set of flexHEG devices.

A more sophisticated approach to such guarantees would be **verifiable evaluations**, i.e. verification that a given model has not exceeded some score on some automated test of model capabilities, or alternatively has exceeded some score on a test of model safety. Such verifiable test results could potentially be produced without sharing the details of the model or its behavior. Verifiable evaluations could even be implemented as a rule: FlexHEG devices could require that capability evaluations are performed regularly during training, and refuse to continue training if a model exceeds a capability threshold. However, making evaluations robust to efforts to train models to underperform on evaluations is a very difficult open problem; flexHEGs could plausibly be used to verify that no such subversive training has taken place, but this is also very difficult to do while preserving privacy.

As our understanding of these risks improves, we may come to understand that models trained in certain ways using certain architectures are much less risky. FlexHEGs could enable **verification of architectures and training techniques**, allowing states to verify to each other that they are using these safer approaches. FlexHEG devices could even be configured to automatically block the training of particularly risky types of models beyond some low compute limit.

# Unsafe Weaponization of AI

Even momentary incidents of loss of control or other misbehavior may be catastrophic in a military context. Nonetheless, states may feel forced to rapidly deploy frontier AI capabilities in their military systems in order to keep up with their rivals [12].

FlexHEGs could alleviate these issues in several ways:
- **Controlled deployment:** FlexHEG devices could be configured to verifiably enforce a rule that they will only allow powerful models trained on them to be deployed to specifically approved other flexHEG devices, possibly with some required deployment-time safeguards. This could allow states to provide credible guarantees that their civilian AI systems are only being deployed in known civilian datacenters, and thus not secretly being deployed in military systems.
- **Verified claims about military systems:** More speculatively, FlexHEGs, if integrated into military AI systems such as drones, could verify claims about the configuration of those military AI systems, such as what kinds of behavior constraints are programmed into the system. However, ensuring and verifying that the flexHEG chip is integrated into the system in such a way that it cannot be swapped out in the event of a conflict is a very difficult problem.
- **Verified claims about model training:** If militaries use flexHEG-equipped data centers to train their own general-purpose models, they could make verifiable claims about what techniques were or were not used to make those systems less risky, similarly to the civilian case discussed in the preceding section. In some cases, militaries may also have an interest in verifying claims about their systems' capabilities, to build confidence that a given system is not capable enough to be a threat to others.

# Threats to Balance of Power and Strategic Stability

AI has the potential to drive significant economic and military transformations, which could threaten the existing balance of power. Some have even proposed that AI could be used to accelerate AI development itself, leading to a dramatic feedback loop that could allow a leading AI developer state to quickly develop a massive technological advantage over their rivals. In addition to exacerbating the risks discussed above, this dynamic could motivate rival states to take destabilizing military action in order to preempt such concentration of power [3], [13], [14], [15].

These competitive dynamics could potentially be resolved in many ways.
- **Controlled deployment** could allow a leading AI developer's models to be deployed on particular flexHEG devices controlled by another state, thus giving the other state irrevocable private black box access to the model, while also giving the developer confidence that the other state cannot modify or further proliferate the model, or remove safeguards.
- **Verified claims about model capabilities:** In some cases, FlexHEGs could allow states to verify that AI systems they are developing do not threaten the security of other states, e.g. because the systems lack particular capabilities. However, as mentioned above, adversarially robust evaluations remain a difficult open problem, and it will likely be difficult to articulate an exhaustive list of concerning capabilities in advance.
- **Multilateral control over AI development:** If other measures are deemed insufficient to defuse the situation, FlexHEG devices could even be configured to require multiple states' approval for certain development actions, such as large new training runs.

## A Sketch of A Path to FlexHEGs

FlexHEGs, particularly as applied to international governance, are quite a novel proposal. However, it appears that the key components could be developed in the coming years, motivated in large part by two key forces: Demand for improved model security, and US export controls. This section will sketch a hypothetical sequence of events that could lead from these forces to an eventual full implementation of flexHEG technology. As part of this, we will sketch some related governance measures that would need to be adopted along the way; these measures will be discussed in more depth in Part III of this series.

Demand for improved security for models and data has already driven the implementation of confidential computing, which provides hardware-level encryption and some verification capabilities. Current GPU confidential computing implementations do not yet support e.g. multi-GPU workloads, but Nvidia has suggested that they will add that support to the next generation [16]. Amazon is also working on adding custom network interface cards (NICs) to their data centers that would allow all between-server traffic to be fully encrypted [17], [18, p. 2429].

Model security is increasingly also becoming a matter of export control policy: In January 2025 the United States announced their AI Diffusion Framework for export controls, that extended controls on chips, and added new controls on models. As part of the framework, starting in 2026 data centers built by "validated end users" for the purposes of deploying export-controlled frontier models in "tier 2" countries need to implement extensive security measures, including ensuring that model weights can only be accessed through a "narrow, well-defined API" to prevent them from being stolen, even by rogue employees with physical access [19].

Security technologies such as confidential computing often include "attestation", i.e. verification, capabilities. These are intended to allow devices to verify their security properties to other devices, but could be used much more broadly. There are already private-sector efforts to build on confidential computing technology to create verification protocols to allow end users to have more trust that, e.g., a fine tuned model has not had a backdoor added in [20], [21]. This kind of verification may also prove useful for demonstrating export control compliance by demonstrating that a set of chips has only been used for the expected, pre-reported end use, and remain where they are expected to be.

With some guidance and foresight by the companies and governments involved, one could imagine these converging demands for improved security resulting in a design that implements much of the general-purpose verification capabilities discussed in this report. This design might even be standardized. In particular, these early systems might make use of a combination of existing confidential computing features, with a custom network interface controller playing the role of the guarantee processor for an entire server or small group of servers; see discussion of "Repurposing the NIC" in Part II.

Physical security of these early systems might be minimal, or might rely heavily on simple solutions such as locks, tamper-evident seals, and simple sensors and security cameras. As a makeshift self-disablement mechanism, the system could wipe its memory and optionally brick itself by wiping its boot ROM if the sensors detect an attempt at unauthorized physical access.

These early real-world deployments would provide valuable data about the design's security and reliability, allowing major flaws to be ironed out, and trust to be built.

## From Best Practice to Requirement

Once a design for high-security, verifiable compute exists, governments could mandate it for certain applications. For example, in the context of the US export controls, the different types of "validated end users" might be required to implement these measures in their data centers in "tier 2" countries, particularly those data centers handling models subject to export controls. This could allow wider deployments to less secure jurisdictions, opening up new markets and supporting benefit sharing [22], [23].

Governments and companies could also use this approach for their own high-security data centers.

At this stage, the supply chain for these components might be subjected to greater scrutiny, and the components would hopefully be thoroughly red-teamed.

## Transition to International Governance

Later, major governments may become convinced of a need to coordinate to regulate AI development to prevent the escalation of international instability [6], or, possibly as a result of some significant AI-related incident, to prevent future incidents. This could be, for example, a notable instance of malicious use, a finding that a new model has potentially dangerous superhuman capabilities in the cyber domain, or even a serious attempt by a frontier model to escape human control.

Before regulating their own AI industries, major governments would want to have confidence that other major governments are also acting similarly responsibly. This could involve, for example, international verification that no new training runs above some FLOP threshold are currently happening without international approval.

At this point, these existing security and verification systems, already found in most data centers in "tier 2" countries, might be deployed much more widely to enable moderately comprehensive verification. This may begin with the data centers being used by frontier labs, but other data centers would soon need to be included to manage concerns about labs conducting secret side projects at other facilities.

This new system would of course not be entirely trustworthy, because e.g. the production of the chips has not been subject to international scrutiny. However, previous testing would hopefully have built up enough trust to allow the system to be relied upon as a temporary solution in the absence of better options.

Substantial efforts could now begin to build further trust in the existing solution, and begin developing more permanent solutions. An international body would likely need to be set up by the key states involved to coordinate and oversee all of this.

There would need to be an effort to establish international trust in the current design, by sharing the design documents and source code, and reviewing each design decision in detail with experts on the other side. Any suspicious, unjustifiable design choices would be investigated, and any design flaws would be mitigated and added to the list of things to fix in the next generation.

At this point, some of the existing components could be sampled for rigorous, destructive reverse-engineering to check that they match the design documents provided by the manufacturer.

The latest software-engineering AIs could be tasked with developing new, sophisticated frameworks for flexibly verifying complex, privacy-preserving claims about workloads, with at least as much compute spent on having yet more AIs scrutinizing the frameworks.

To ensure that no one is circumventing the verification requirements by building secret data centers, all relevant chip fabs would likely need to be subjected to monitoring, and all powerful AI chips would likely be registered in a database operated by the international oversight body[3].

The facilities for producing the key components would likely need to be reviewed, separated from other production, and subjected to extensive oversight. Based on findings about the current commercial design, a new internationally approved design could be quickly drafted and subjected to extensive testing and verification.

---

[3] See additional discussion in Part III.

Flaws in the early verification systems might still occasionally be discovered, but the most important data centers should at this point be subjected to many layers of protections, including physical inspections, whistleblower incentives, and conventional espionage, making it difficult for frontier developers to get away with exploiting the flaws.

When extremely powerful AI systems eventually become more commonplace, they would be deployed on tightly secure flexHEG servers, using custom secure enclosures, with the old flaws ironed out. The models can thus be accessed privately, yet safely.

Occasionally rogue actors might attempt to stockpile non-flexHEG AI chips or consumer GPUs [24], but AI supported intelligence agencies could discover and stop them before they could do real harm.

# FlexHEGs as a Comprehensive Governance Framework

If flexHEGs became widely deployed, it would be possible to use them to create a comprehensive international governance framework for AI.[4] Such a framework could cover all or nearly all relevant data center compute, and flexibly address a wide range of issues as they emerge, without needing to renegotiate a new treaty in response to each new problem.

This framework could be strictly verification-based, i.e. based on comprehensive verification of compliance, without any formal enforcement mechanisms. There would be nothing directly preventing violations of norms, but verification would still enable a collaborative equilibrium through transparency and accountability.

Alternatively, in a ruleset-based framework flexHEG devices could be configured to automatically enforce a multilaterally agreed-upon ruleset. The ruleset could be changed via firmware and software updates, but these updates would need to be signed by some key states, or some large majority of all participating states. Such a ruleset could be complemented by verification of claims that would be difficult to formalize as enforceable rules. This approach offers several advantages:
- Even the most powerful states could not easily unilaterally abandon the agreement. This would extend the agreement's guarantees into the future, and improve international stability by e.g. allowing smaller states to trust more powerful states' guarantees about how they will use AI technology.
- Even rogue actors could not violate rules. This would reduce the need to closely control which actors can access flexHEG compute.

---

[4] In addition to the flexHEG devices themselves, this framework would include other efforts to track compute to prevent the flexHEG-based mechanisms from being circumvented. This was briefly sketched in the preceding section, and is discussed in more depth in Part III.

However, states capable of chip production could still gradually exit the agreement unilaterally by resuming production of non-flexHEG chips, so even a ruleset-based agreement could not bind all participants permanently.

Both verification-based and ruleset-based frameworks could likely be stable: As long as the expected geopolitical advantage to be gained from defecting from the agreement is smaller than the expected loss of security caused by the unraveling of the agreement, it would be rational for the major players not to defect. We develop a simple game theoretic model of this in Appendix B of Part III, which suggests that if defections could be detected quickly enough that the advantage to be gained from defection would be limited (≤75% chance of winning), the agreement would be stable as long as the participants believe that developing AI without coordination would be at least moderately dangerous (≥10% chance of catastrophe) and they do not see winning a race for dominance as massively better outcome than maintaining the existing balance of power through an agreement (≥1.5 times better).

Finally, we must acknowledge that a comprehensive, global regime aimed at controlling the use of compute would have significant potential for abuse. However, the potential for such abuse could be greatly reduced through the right design choices. Most importantly, ruleset-based agreements could be arranged to require all rules to be approved by several different major states. This would make it difficult for the rules to be used to advance any one state's particular political goals, and would help ensure that the rules are focused on matters where there is consensus that there is a genuine international security concern.

On the other hand, flexHEGs configured for verification only, with no enforcement mechanisms, would not give states any coercive power that they did not already have. Even comprehensive surveillance capabilities are already feasible for determined states to attain.

It is also crucial to remember that this entire proposal is premised on the assumption that AI technology could be as dangerous as nuclear technology. A comprehensive agreement is very unlikely to be realized unless major states become convinced of this. If they do, they will likely attempt to implement some form of oversight regardless. FlexHEGs could enable that oversight to be much more privacy-preserving, transparent, and (optionally) multilateral than it would be otherwise.

# Limitations of FlexHEG Mechanisms

While flexHEGs would be very powerful if fully implemented, they still have some important limitations.

## FlexHEGs are Technically Challenging to Implement

A key drawback of the flexHEG proposal is that it is technically quite ambitious. Achieving a very high standard of physical security and developing sophisticated verification protocols could take several years. As a result, these efforts may need to begin several years before verification capabilities of this caliber are actually needed, requiring foresight from the governments and companies involved.

Even after the ecosystem itself has been set up, continuously adjusting verification protocols to adapt to changes in the underlying technology may also prove to be expensive and slow, though we may reasonably hope that progress in AI programming assistance will reduce this cost.

## FlexHEGs Could Only Govern the Frontier of AI

Like other compute governance approaches [25], FlexHEGs will likely only be able to effectively govern the most compute-intensive frontier of AI development.[5] Various motivated actors will always be able to access at least some concentrations of compute that are not flexHEG-equipped and have fallen outside other oversight mechanisms. As a result, flexHEG will not fully obviate the need for e.g. conventional intelligence efforts to identify terrorist organizations that might attempt to use AI maliciously.

Notably, one of the unique advantages of flexHEGs compared to other forms of compute governance is that they plausibly *could* efficiently govern even widespread deployment of frontier models if flexHEG devices are configured to guarantee that frontier models trained on flexHEG devices cannot leave the flexHEG ecosystem, and can only be deployed in specific ways on approved flexHEG devices.

## Technical Verification Has Limits

Certain important aspects of AI development and use may not be readily verifiable on-chip. Most importantly, various notions of "misuse" and malicious use of AI ultimately depend on what is done with the result of some computation, which the flexHEG device itself cannot know. For example, if flexHEG devices are used to look for vulnerabilities in some computer system, the devices themselves have no way of knowing whether the intent is to repair or to exploit those vulnerabilities.

---

[5] See further discussion in Part III.

FlexHEGs can only distinguish between acceptable and unacceptable uses insofar as the computation itself is significantly, technically distinct between these two cases. This means that the flexHEG proposal is partly based on the hope that we will be able to develop a sufficiently advanced understanding of AI systems and their applications to be able to articulate such distinctions.

### Other Governance Approaches Can Complement FlexHEGs

Using FlexHEGs alongside other approaches and information sources could compensate for many of these limitations. This could include on-site inspections, whistleblower rewards, national technical means of verification, financial intelligence, and other intelligence capabilities [26], [27]. Many of these could be implemented more quickly, and thus should be kept in the portfolio in case flexHEGs cannot be implemented quickly enough. Intelligence capabilities could also help catch types of malicious use that are too small-scale for flexHEGs to cover. These additional types of information could also provide the context to distinguish malicious from helpful, offensive from defensive, and military from civilian.

On the other hand, these less technically sophisticated mechanisms may also make flexHEGs entirely unnecessary if there is sufficient trust between the most relevant states that verification methods need not be extremely secure or privacy-preserving.

## Recommendations

To many readers, flexHEG may appear excessive or overly ambitious as a solution to the current issues posed by AI. However, we emphasize that flexHEG specifically addresses potential future risks associated with advanced AI systems. Many experts predict that very significant novel risks could emerge in the coming years. Given the considerable time likely needed for implementation, development may need to begin soon in order for the technology to be ready in time.

Other readers may conversely think that the need for solutions like this is so urgent that this proposal is unworkable because it would take too long to implement. One of these skeptical perspectives may indeed prove to be correct. We nonetheless believe that flexHEGs or related technology are likely enough to prove to be valuable in one way or another, that it is worth investing in R&D to begin developing this technology.

In addition to the several applications of flexHEGs that were discussed above, this R&D could work prove useful in a range of different ways:

- General improvements to hardware and software security, for AI compute and otherwise, are obviously valuable in innumerable ways, including helping secure AI model weights [28, p. 91].
- Less secure and trustworthy implementations of some aspects of flexHEG capabilities could be useful for facilitating more privacy-preserving domestic regulation.
- Even if only small, prototype quantities of internationally verifiable compute could be produced, this could enable valuable verification exercises such as internationally verifiable evaluations of models.
- Even if a given verification technology is not fully internationally trusted, it can still help build some additional confidence regarding states' intentions and behavior, together with other signals.
- Even if the technology is not ready before extremely powerful AI is realized and many of the problems discussed above have already become acute and possibly been solved, improved verification technology would likely still prove to be ever more important in a world transformed by AI.

## Recommended Areas of Technical Research

While an ideal flexHEG design would be designed from scratch with reliability and trust in mind, realistically designs should aim to make use of existing technology wherever possible, and to be retrofittable on existing AI chip and server designs. Initial R&D efforts could focus on:

**Developing an open source IP block** that could be used as the guarantee processor, either as a separate chip or integrated into a larger chip design.[6] This could include extending existing designs such as OpenTitan [29] or Caliptra [30]. More ambitiously, companies should be encouraged to open source their existing modules, such as Nvidia's GPU System Processor [31].

**Developing verification protocols** based on the verification capabilities of existing confidential computing features found on GPUs and CPUs.[7] (These are more commonly referred to as *attestation* capabilities.) Specific early targets for verification could include:
- Claims about how a given model was trained, starting with the simple claim of how many FLOP were used.
- Verifiable, mutually privacy-preserving evaluations that keep model properties secret from the evaluator, while keeping the content of the evaluation secret from the developer [32].
- Controlled deployment: Ways to use confidential computing to verifiably deploy models to GPUs in a way that guarantees the confidentiality of the model weights and the integrity of all model safeguards.

---

[6] See "Project 2" in Part II for more detail.
[7] See "Project 3" and "Project 4" in Part II.

**Developing methods to make already-deployed accelerators tamper-evident** using existing technologies like specialized seals and camera systems.[8]

**Analyzing existing processors within accelerator components** to determine whether they have appropriate capabilities and features, and are secure enough, such that they could be repurposed via firmware updates to function as a guarantee processor without hardware modifications.[9]

**Designing frameworks to accurately verify and log the operations** performed by accelerators.[10]

**Developing robust algorithms to locally check** whether accelerator usage is compliant with guarantees.[11]

## Policy recommendations

In addition to directly funding R&D efforts, if policymakers want to accelerate the development and deployment of flexHEGs or similar technical guarantees, they could set policies that create a demand for the technology [33]. This could include:
1. Including requirements for hardware security features as part of export control policies such as the US AI diffusion framework [19], [34].
    a. Location verification could be required to ensure that chips have not moved [11], [19][12].
    b. Hardware security and encryption features could be required when deploying models to untrusted jurisdictions [19][13].
    c. Tamper-response mechanisms could be required as a precaution against an untrusted jurisdiction seizing chips [19][14]. This could include wiping memory and storage and even "bricking" the chip by wiping key data required for the system to boot.
2. Requirements for tamper protections and attestation capabilities could be added to security standards for data centers, such as FedRAMP.
3. Tamper protection and attestation capabilities could be added as requirements for chips procured for high-security government applications, if they would not significantly hamper that procurement.

---

[8] See "Project 1" in Part II.
[9] See "Firmware Modification" in Part II.
[10] See "Transparent Logging and Auditing", "Project 3", and "Project 4" in Part II.
[11] See "General-Purpose Workload Guarantees" and "Project 4" in Part II.
[12] § 748, Supplement No., 8(B)(11)
[13] § 748, Supplement No. 10, (14)(d)
[14] § 748, Supplement No. 10, (14)(c)

In many cases, a bottleneck to the above is that there are no applicable existing standards for the relevant components, such as tamper protections suitable for AI chips. The development of such standards is another measure through which policymakers could support the development of flexHEG technology.

## Conclusion

The development of flexHEG mechanisms presents significant technical challenges, but the potential benefits for international security justify the required investment. Even partial implementations of the technology could create valuable options for addressing governance challenges as they emerge.

What makes flexHEGs particularly valuable is not just their technical capabilities, but their ability to align technical solutions with geopolitical realities. By enabling states to make and verify commitments while preserving privacy and sovereignty, flexHEGs offer a pragmatic approach that respects the competitive dynamics of international relations while still allowing for crucial coordination on matters of shared risk.

The history of technology governance shows that frameworks established early can shape development trajectories for decades. Whether flexHEGs eventually become part of a comprehensive governance regime or serve more limited verification purposes, beginning development now ensures these options remain available as AI capabilities advance, potentially providing an essential component for maintaining both technological progress and international stability.

# Bibliography

[1] J. Petrie, O. Aarne, N. Ammann, and D. "davidad"Dalrymple, "Interim Report: Mechanisms for Flexible Hardware-Enabled Guarantees," Aug. 2024. [Online]. Available: https://yoshuabengio.org/wp-content/uploads/2024/09/FlexHEG-Interim-Report_2024.pdf

[2] "Statement on AI Risk." Center for AI Safety, 2023. [Online]. Available: https://www.safe.ai/work/statement-on-ai-risk

[3] J. Mitre and J. B. Predd, "Artificial General Intelligence's Five Hard National Security Problems," RAND Corporation, Feb. 2025. [Online]. Available: https://www.rand.org/pubs/perspectives/PEA3691-4.html

[4] "IDAIS-Venice Statement," Sep. 2024. [Online]. Available: https://idais.ai/dialogue/idais-venice/

[5] "Frontier AI Safety Commitments, AI Seoul Summit 2024," Department for Science, Innovation and Technology. [Online]. Available: https://www.gov.uk/government/publications/frontier-ai-safety-commitments-ai-seoul-summit-2024/frontier-ai-safety-commitments-ai-seoul-summit-2024

[6] D. Hendrycks, E. Schmidt, and A. Wang, "Superintelligence Strategy: Expert Version," Mar. 07, 2025, *arXiv*: arXiv:2503.05628. doi: 10.48550/arXiv.2503.05628.

[7] A. J. Coe and J. Vaynman, "Why Arms Control Is So Rare," *Am. Polit. Sci. Rev.*, vol. 114, no. 2, pp. 342–355, May 2020, doi: 10.1017/S000305541900073X.

[8] E. Stafford, R. F. Trager, and A. Dafoe, "Safety Not Guaranteed: International Races for Risky Technologies," Centre for the Governance of AI, Nov. 2022. [Online]. Available: https://www.governance.ai/research-paper/safety-not-guaranteed-international-strategic-dynamics-of-risky-technology-races

[9] Confidential Computing Consortium, "A Technical Analysis of Confidential Computing," Nov. 2022, [Online]. Available: https://confidentialcomputing.io/wp-content/uploads/sites/10/2023/03/CCC-A-Technical-Analysis-of-Confidential-Computing-v1.3_unlocked.pdf

[10] Y. Bengio *et al.*, "International AI Safety Report," DSIT 2025/001, 2025. [Online]. Available: https://www.gov.uk/government/publications/international-ai-safety-report-2025

[11] A. Brass and O. Aarne, "Location Verification for AI Chips," Institute for AI Policy and Strategy, Apr. 2024. [Online]. Available: https://www.iaps.ai/research/location-verification-for-ai-chips

[12] J. Allen and D. West, "Op-ed: Hyperwar is coming. America needs to bring AI into the fight to win — with caution," *CNBC*, Jul. 12, 2020. [Online]. Available: https://www.cnbc.com/2020/07/12/why-america-needs-to-bring-ai-into-the-upcoming-hyperwar-to-win.html

[13] L. Aschenbrenner, "Situational Awareness: The Decade Ahead," Jun. 2024. [Online]. Available: https://situational-awareness.ai/

[14] H. A. Kissinger, E. Schmidt, and C. Mundie, *Genesis: Artificial Intelligence, Hope, and the Human Spirit*. Little, Brown and Company, 2024.

[15] C. Katzke and G. Futerman, "The Manhattan Trap: Why a Race to Artificial Superintelligence is Self-Defeating," Dec. 22, 2024, *arXiv*: arXiv:2501.14749. doi: 10.48550/arXiv.2501.14749.

[16] "NVIDIA Blackwell Architecture Technical Overview," NVIDIA, 2024. [Online]. Available: https://resources.nvidia.com/en-us-blackwell-architecture

[17] D. Patel, W. Chu, C. Tseng, M. Xie, J. Eliahou Ontiveros, and D. Nishball, "GB200 Hardware Architecture – Component Supply Chain & BOM," *SemiAnalysis*, Jul. 17, 2024. [Online]. Available: https://semianalysis.com/2024/07/17/gb200-hardware-architecture-and-component/

[18] "Amazon Elastic Compute Cloud - User Guide." Amazon Web Services, 2024. [Online]. Available: https://docs.aws.amazon.com/pdfs/AWSEC2/latest/UserGuide/ec2-ug.pdf#data-protection

[19] Bureau of Industry and Security, "Framework for Artificial Intelligence Diffusion," Federal Register, 2025–00636 (90 FR 4544), Jan. 2025. [Online]. Available: https://www.federalregister.gov/documents/2025/01/15/2025-00636/framework-for-artificial-intelligence-diffusion

[20] "Verifiable Compute White Paper," EQTY Lab, 2024. [Online]. Available: https://www.eqtylab.io/verifiable-compute-white-paper

[21] A. Aguirre and R. Millet, "Verifiable Training of AI Models," *Future of Life Institute*, Jul. 23, 2024. [Online]. Available: https://futureoflife.org/ai/verifiable-training-of-ai-models/

[22] C. O'Keefe, "Chips for Peace: How the U.S. and Its Allies Can Lead on Safe and Beneficial AI," *Lawfare*, Jul. 10, 2024. [Online]. Available: https://www.lawfaremedia.org/article/chips-for-peace--how-the-u.s.-and-its-allies-can-lead-on-safe-and-beneficial-ai

[23] C. Dennis *et al.*, "Options and Motivations for International AI Benefit Sharing," Centre for the Governance of AI, Jan. 2025. [Online]. Available: https://www.governance.ai/research-paper/options-and-motivations-for-international-ai-benefit-sharing

[24] E. Grunewald, "Are Consumer GPUs a Problem for US Export Controls?," Institute for AI Policy and Strategy, May 2024. [Online]. Available: https://www.iaps.ai/research/are-consumer-gpus-a-problem-for-us-export-controls

[25] G. Sastry *et al.*, "Computing Power and the Governance of Artificial Intelligence," Feb. 13, 2024, *arXiv*: arXiv:2402.08797. doi: 10.48550/arXiv.2402.08797.

[26] A. Scher and L. Thiergart, "Mechanisms to Verify International Agreements About AI Development," Machine Intelligence Research Institute, Nov. 2024.

[27] A. R. Wasil, T. Reed, J. W. Miller, and P. Barnett, "Verification methods for international AI agreements," Aug. 28, 2024, *arXiv*: arXiv:2408.16074. [Online]. Available: http://arxiv.org/abs/2408.16074

[28] S. Nevo, D. Lahav, A. Karpur, Y. Bar-On, H. A. Bradley, and J. Alstott, "Securing AI Model Weights: Preventing Theft and Misuse of Frontier Models," RAND Corporation, May 2024. [Online]. Available: https://www.rand.org/pubs/research_reports/RRA2849-1.html

[29] A. Meza, F. Restuccia, J. Oberg, D. Rizzo, and R. Kastner, "Security Verification of the OpenTitan Hardware Root of Trust," *IEEE Secur. Priv.*, vol. 21, no. 3, pp. 27–36, May 2023, doi: 10.1109/MSEC.2023.3251954.

[30] CHIPS Alliance, "Caliptra: A Datacenter System on a Chip (SoC) Root of Trust (RoT)," GitHub. [Online]. Available: spec.caliptra.io

[31] AleksandarK, "NVIDIA Unlocks GPU System Processor (GSP) for Improved System Performance," TechPowerUp. [Online]. Available: https://www.techpowerup.com/291088/nvidia-unlocks-gpu-system-processor-gsp-for-improved-system-performance

[32] A. Trask *et al.*, "Secure Enclaves for AI Evaluation," OpenMined Blog. [Online]. Available: https://blog.openmined.org/secure-enclaves-for-ai-evaluation/

[33] T. Fist, T. Burga, and V. Chilukuri, "Technology to Secure the AI Chip Supply Chain: A

[33] Working Paper," Center for a New American Security, Dec. 2024. [Online]. Available: https://www.cnas.org/publications/reports/technology-to-secure-the-ai-chip-supply-chain-a-primer

[34] L. Heim, "Understanding the Artificial Intelligence Diffusion Framework: Can Export Controls Create a U.S.-Led Global Artificial Intelligence Ecosystem?," RAND Corporation, Jan. 2025. doi: 10.7249/PEA3776-1.

[35] Y. Bai *et al.*, "Constitutional AI: Harmlessness from AI Feedback," Dec. 15, 2022, *arXiv*: arXiv:2212.08073. doi: 10.48550/arXiv.2212.08073.

[36] "Introducing the Model Spec," *OpenAI*, May 08, 2024. [Online]. Available: https://openai.com/index/introducing-the-model-spec/

[37] R. Greenblatt, B. Shlegeris, K. Sachan, and F. Roger, "AI Control: Improving Safety Despite Intentional Subversion," Jul. 23, 2024, *arXiv*: arXiv:2312.06942. doi: 10.48550/arXiv.2312.06942.

[38] O. Aarne, T. Fist, and C. Withers, "Secure, Governable Chips," Center for a New American Security, Jan. 2024. [Online]. Available: https://www.cnas.org/publications/reports/secure-governable-chips

[39] G. Kulp *et al.*, "Hardware-Enabled Governance Mechanisms: Developing Technical Solutions to Exempt Items Otherwise Classified Under Export Control Classification Numbers 3A090 and 4A090," RAND Corporation, Jan. 2024. [Online]. Available: https://www.rand.org/pubs/working_papers/WRA3056-1.html

[40] M. Wortsman *et al.*, "Model soups: averaging weights of multiple fine-tuned models improves accuracy without increasing inference time," in *Proceedings of the 39th International Conference on Machine Learning*, PMLR, Jun. 2022, pp. 23965–23998. [Online]. Available: https://proceedings.mlr.press/v162/wortsman22a.html

# Appendix A: The Original FlexHEG Vision

The original vision for flexHEGs, developed by David 'davidad' Dalrymple, was a mechanism that would implement a transparent, updateable, multilaterally agreed-upon ruleset fully automatically on-device, while being free of any software bugs and impervious to physical tampering. This would mean that there would be no need for the device to send any information back to any authority, and no need for any authority to even know who owns each device or where it is. All of this would protect individuals from privacy violations and abuses of power, while allowing governments and everyone else to trust that the devices would enforce the rules and keep the world secure.

A critical element of the original vision was that the ruleset would contain restrictions on future rulesets, ensuring that future updates could not undermine these privacy and security guarantees by e.g. adding requirements for devices to report information back to authorities.

## How We Expanded the Definition

This original vision could be thought of as the pinnacle of flexHEGs. However, in the process of writing this report, we came to believe that, at least for the initial iterations of flexHEG technology, key aspects of this vision are likely technically and politically infeasible to achieve. Therefore this report broadens this vision by relaxing some of the original vision's assumptions.

This report's definition of flexHEGs retains the goal that they should be transparent and updateable, and should not rely on "phoning home" to implement any of its basic functions. However, we do somewhat diverge from the original vision in that we extensively discuss verification approaches that involve users choosing to send at least minimal information to others. This choice could be fully free, or compelled by e.g. a government, but in almost all cases it would not – and in many configurations could not – be compelled by the flexHEG device itself.

Second, we treat on-device enforced rulesets – multilateral or otherwise – as merely one of the ways to configure flexHEGs. Using flexHEGs purely to verify compliance with rules is discussed as a major alternative, along with many variations falling somewhere in between these two.

## Feasibility Problems With the Original Vision

The original vision appears technically or politically infeasible for several reasons:

Checking compliance with rules fully automatically on-device appears unlikely to be feasible, for several reasons. For example, malicious intent is not a technical property observable on-chip, and many safety properties of frontier systems likely cannot, at least initially, be defined precisely enough to be evaluated automatically on-device, in an adversarial setting.[15]

If relevant policy goals can't be checked on-device, it becomes necessary to share at least some information with some authorities. This compromises the original vision of perfect privacy, where the authorities have no idea where the chips are and what they are doing.

Relatedly, even if it were possible to check all the relevant claims fully on-device, the original vision requires tremendous trust that the mechanisms for doing so, as well as the update mechanism and the device's physical tamper protections, are all essentially flawless. It is implausible that such trust could be justified for a technology before it has seen widespread real-world deployment. Therefore e.g. occasional physical inspections of flexHEG devices are likely a practical necessity for early iterations of the technology. Physical inspections in turn require a registry of device owners and locations, further undermining original vision of absolute privacy.

Even if these technical difficulties could be solved, a binding, internationally agreed-upon ruleset could be perceived as violating states' sovereignty, and could also be less stable than a primarily verification-based approach[16].

A key technical requirement of the original vision is the ability to restrict the content of future updates. Without this, there is no guarantee that flexHEGs would not be used in a way that violates aspects of the original vision, such as adding a rule that *de facto* requires "phoning home". This is most likely not feasible to implement, at least on early iterations of flexHEGs.[17]

Restriction of future updates essentially requires defining a formal language for describing rules, precisely defining relevant notions such as "phoning home" or "adding a backdoor" in terms of that language. You then additionally need an non-updateable bottom layer of firmware that enforces these rules. This may be possible in the future, but it is quite difficult to get the firmware perfectly correct initially, and to define the rule language to be general enough to allow sufficient flexibility, while being simple enough that we can successfully reason about what would constitute a "backdoor", "kill switch", or "phoning home".

Even if restriction of future updates were feasible, there would be little guarantee that those with the power to set the rules would in fact tie their own hands in this way.

Overall, it appears likely that the best way to achieve the spirit of the original vision, such as not having the chip force the user to phone home or reveal their location, would be to configure the flexHEG devices for verification (and perhaps binding commitments) only, with no mechanism for forced updates.[18]

---

[15] See Part III.
[16] See Part III.
[17] See "Guarantee Update Process" in Part II.

# Appendix B: Governance Mechanisms Enabled by FlexHEGs

This appendix will provide more detailed sketches of how flexHEGs could be used to implement a wide range of guarantees of different types. It is intended to act as a reference for possible guarantee mechanisms, but should not be thought of as an exhaustive list, given the flexibility of flexHEGs.

We will first discuss how to categorize different kinds of mechanisms, before moving on to discussing how various specific mechanisms in each of these categories might be implemented.

These mechanisms, and the descriptions of their working, should be thought of as rough sketches. We are not entirely confident that all of these would be practically feasible, much less desirable, to implement.

The focus of this appendix will be on explaining how different mechanisms would work on a conceptual level. More detailed discussion of technical implementation details, requirements and feasibility will be deferred to Part II. On the other hand, detailed discussion of how these mechanisms could be used to solve real-world governance problems will be deferred to Part III.

## Categories of Guarantees

- **Verification mechanisms** are mechanisms that allow the user of a chip (the **prover**) to make verifiable claims about what they have done or are doing with a flexHEG device to some other party (the **verifier**).
    - Verification mechanisms can be based on verifiable sharing of detailed information, such as detailed logs of an entire training run, including snapshots of weights during training.
    - Alternatively, verification mechanisms can be **privacy-preserving** if a flexHEG device or set of devices can verify the details of the claim locally, and can thus only share the absolute minimum amount of information. A privacy-preserving verifiable claim might be of the form "the use of these devices up to X point in time has always been in compliance with Y ruleset", with no additional information shared with the verifier.
- **Rules** are guarantees that a given flexHEG device cannot be used to perform workloads that would violate a given ruleset. Typically a ruleset would initially be configured at the factory, and could be updated later with the right authorizations.

---

[18] See Part III.

- **Binding commitments** are rules that the chip owner has voluntarily subjected their chips to, but in such a way that the user themselves cannot remove the rules for some period of time, and other parties can verify that this is the case.
- **Discretionary enforcement mechanisms** are mechanisms that require some authority's approval for the device to perform some or any actions.
- **Baseline rulesets** are simple, relatively restrictive rulesets that a flexHEG device could be configured to revert to if its operating license or full ruleset expire.

# Verification

## Verifying Claims about Workloads

Put in general terms, various computing workloads performed on AI chips can be thought of as taking some data as inputs and processing that data in some way to produce different data. Generic data analysis typically involves taking some data, filtering and transforming it in some way, and calculating statistics over the resulting "cleaned" data. Training a machine learning model typically looks like taking randomly initialized weights and a dataset and processing them together to modify the weights to perform better on some task. Deploying a model involves taking some input data from a user, and processing it using the model weights to produce an output. Outside of AI, running e.g. a simulation will typically involve taking some starting state as data and processing it according to some set of rules to obtain the next state of the simulation, and so on.

Guarantee processors could be used to verify various types of workloads across large numbers of flexHEG devices by producing verifiable "receipts" of what data were processed and how, resulting in a description of how some final result (such as a trained model, or a model output) was produced. Guarantee processors could aggregate and analyze these receipts and produce higher-level receipts that make more abstract claims about the workload. For example, receipts that "devices *A*, *B*, and *C* contributed *X*, *Y*, and *Z* FLOP to intermediate result *K*" could be aggregated to a receipt that "intermediate result *K* was produced using a total of *X+Y+Z* FLOP", which could be further simplified to "*K* was produced with less than *L* FLOP". This aggregation allows higher-level receipts to include only the minimal amount of information that needs to be reported, keeping details private.

This kind of privacy-preserving verification requires that the mapping from the low-level information contained in the receipts to the high-level claims that the verifier is interested in can be done automatically on-device. In some cases this may not be possible, for example if the claim to be verified cannot be stated technically precisely. This may be particularly likely for some notions of a given trained model being "safe"[19].

If automated mapping is not possible, a less privacy-preserving option would be to share the low-level receipts themselves for manual analysis by the verifier. This could still be somewhat privacy-preserving: For example, a representative of the verifier could access the data in a controlled environment where they cannot take away any information beyond that which they can remember, and the prover can oversee the representative's activities to ensure that they do not attempt to extract additional, unnecessary information from the data. In the future, this representative of the verifier could perhaps be an AI agent, enabling much stronger guarantees that the agent will not share any additional information beyond the main conclusion of their analysis.

---

[19] See Part III.

The description given here is simplified and overlooks some open problems, such as how to ensure that additional compute is not implicitly smuggled in, e.g., through inputs that are supposedly training data. However, these appear to be likely solvable; see the discussion of "Multi-Accelerator FLOP Counting" in Part II for more possible solutions.

### Verifying Quantity of Compute Used

To determine the total computing capacity used to train a particular model, each guarantee processor can keep track of how much computation has gone into producing various intermediate results, such as model outputs, activations, gradients, and shards of weights, and pass that information to the next processor in a verifiable "receipt" accompanying the intermediate result itself. These receipts can then be traced back all the way to the randomly initialized starting point and summed to obtain the total number of floating-point operations that contributed to a final set of model weights.

In practice, this requires somewhat detailed tracking of where exactly the FLOP going into a given result came from, to avoid double-counting and other issues. We give a more detailed description of an implementation of "Multi-Accelerator FLOP Counting" in Part II.

### Verifying Data Used

FlexHEGs could also be used to verify claims about which data, or how much data, was used to train a given model.

Verifying upper bounds on the quantity of data could be done similarly to FLOP counting, by counting how much data was used to create each intermediate result.

The "receipts" could also contain hashes of the specific data used, allowing developers to later prove to auditors which specific data were used. This would not necessarily be privacy-preserving, but it may be possible to develop privacy-preserving protocols for doing the audits.

It may also be possible to verify that specific types of training data were not used. A naive solution could include verifying that all of the data used came from specific known, audited datasets.

## Verifying Model Properties and Training Techniques

Verifiable records of the training of a model could also be used to verify claims about the architecture of a model, as well as claims about whether specific training techniques were used.

For example, receipts might in principle be analysed to determine whether a given model was trained using reinforcement learning. While simply determining whether a model was subject to reinforcement learning may not be very useful, hopefully in the future AI research will produce a deeper understanding of how e.g. safety properties depend on specific architectural decisions and training methods used, making this kind of verification more useful.

It may also be possible to verify what goals a given model was trained to follow, but this would currently be somewhat difficult as models may not have precisely defined goals, but it may be possible to verify e.g. that a model was trained in a particular way to follow a particular "constitution" [35] or "model spec" [36]. While this information is in principle possible to derive from information that flexHEG devices can verify, how exactly to implement such verification in practice, particularly in a privacy-preserving way, is still an open problem.

## Verifying Deployments

After a model has been trained, it may also be relevant to verify claims about how it is being deployed. This could be done by producing receipts of inference workloads.

Specific claims of interest to verify about deployments may include:
- Verifying that a given model being deployed has been verified to have been trained or evaluated to be compliant with some norms or regulations.
- Verifying that particular safeguards intended to prevent certain forms of use of the model (e.g. military use or particular forms of malicious use [10]) have been in place.
- Verifying that a deployed model has been subjected to particular automated oversight mechanisms during deployment to protect against misbehavior by the model [37].

## Verifying Negative Claims Through Compute Accounting

Receipts could also be combined to make claims about what has *not* been done by combining receipts describing what has been done. For example, an AI developer may want to prove that they have not trained a model above some FLOP threshold. They could do so by providing privacy-preserving verification of all of the workloads they have run in their data centers that prove that none of those workloads contributed toward such a training run, and that the total FLOP used by the workloads adds up to the total capacity of the data centers. We call this "compute accounting", after Baker et al. (forthcoming). Similarly, someone who operates a large number of AI chips for the purposes of running simulations unrelated to AI could aggregate receipts describing all of their simulation workloads from a particular month to produce a receipt stating "this set of chips was not used for AI workloads during this particular month." If some chips were simply idle or offline, the guarantee processor would be able to verify this as well.

## Verifiable and Privacy-Preserving Evaluations

Many relevant claims about models, such as their capabilities or safety properties, are difficult to precisely verify based on technical claims about how the model was trained. A more useful way to assess a model's properties is to run evaluations, i.e. tests, of the trained model's behavior. FlexHEGs could enable evaluations to be run such that it could be verified that the evaluation is actually being run against the relevant model, without revealing the details of the model.

This could be combined with verification of claims about training to prove that particular automated evaluations were run at regular intervals during training, e.g. to monitor how the model's capabilities developed.

FlexHEGs would also allow evaluations to be mutually privacy-preserving, i.e. the model weights and architecture could be kept hidden from the evaluator, while also keeping the content of the evaluation secret from the model developer. The flexHEG device could guarantee to both parties that the only thing revealed to the other will be the ultimate result of the evaluation. This has already been prototyped using existing confidential computing implementations [32]. Keeping the content of the evaluation secret would be particularly important to protect against the developer training their model to e.g. underperform its true capabilities on specific evaluation tasks, or only behave safely on specific evaluation tasks.

## Location Verification

The locations of FlexHEG devices could be verified by measuring the latency of verification to a trusted landmark, which may itself be a flexHEG device. Location verification like this only requires that a device can reliably attest to its own identity, so full flexHEG capabilities would not be needed, but increased physical protections against the device's identity key being extracted would make location verification by flexHEG devices much more trustworthy than implementations based on existing capabilities.

The verification can be implemented either by having the trusted landmark reach out to the flexHEG device and measure the response time, or by having the flexHEG device itself be the one reaching out, potentially without identifying itself to the trusted landmark. This would allow the flexHEG device to determine its own location without revealing it to the landmark. The device could then make limited, verifiable claims about its location based on this measurement, such as "this device is in an allowable area" without revealing even an imprecise location measurement.

Even if the location measurement is performed by the landmark, this can be relatively privacy-preserving as it is typically very imprecise [11]. Typically the precision would only be at the country level, and the verifying flexHEG device can in principle add delay to its own responses to reduce the precision of the measurement until the measurement only barely confirms that the device is in the expected region. On the other hand, if the trusted landmark is very close to the flexHEG device, the precision of the location measurement can potentially be very high, if desired.

### Configuration Verification

In some cases, it may be desirable to verify the configuration of a set of flexHEG devices. For example, if a user can verify that their flexHEG devices have only been configured into small clusters, with only limited bandwidth between the cluster and the outside world, they might be exempt from expectations to verify in more detail what they are doing on that cluster.

On the other hand, a data center operator may want to verify that all of the devices they have bought for a given data center are actually currently networked together, to prove that none of the devices have been diverted to other facilities.

We give a more detailed description of how configuration verification could be implemented as part of "Encrypted Cluster Formation" in Part II.

### Verifying Claims About Sharing Results

FlexHEG devices could be configured to keep detailed logs of what information they have sent to which other flexHEG chips. This would allow them to be used to verify claims about how some result of some computation has been shared, such as "the weights of this model have only ever been sent to this specific set of flexHEG devices, none of which have not sent those weights outside that set". Verifiable communication like this is generally possible because the data can be encrypted using the specific recipient device's public key such that it cannot be decrypted by any other device.

Combining this with other forms of verification allows verification of claims regarding the entire development and deployment history of a model, for example, that a given model has never been deployed without given safeguards in place.

## Rules

If flexHEG devices can verify a claim in a privacy-preserving way, i.e. if they can check the claim on-device based on receipts they have access to, they can likely also implement automatically enforced rules about such claims. For example, one of the most basic examples of a rule would be a rule limiting AI training workloads to a maximum compute threshold could be enforced by having the guarantee processor block any operation that would result in a training run exceeding this threshold, e.g. as a result of the operation merging or further processing previous intermediate results.

In addition to compute limits, many verifiable claims could straightforwardly be made into rules:
- Any claims about training data, such as total training data quantity, that can be verified on-device, could be made into rules restricting e.g. maximum total training data for particular types of models.
- Any on-device verifiable model properties or training techniques could potentially be restricted on-device.
- Verifiable evaluations could be required to be performed at regular intervals when training particular types of models, and further training could be blocked if the evaluation is not performed, or if the result suggests that the model is e.g. currency unsafe in some way or or exceeds some capability threshold.
- The sharing of models and other computational results could be restricted, for example that models trained with more than X FLOP can never be allowed to be sent to a non-flexHEG device.
- Conditions could be placed on how models can be deployed. For example, models trained with more than X FLOP might only be allowed to be deployed (on flexHEG devices) with particular deployment-time safeguards in place. We call this **controlled deployment**.

The ruleset would be updateable. The details of this update process could be set up in several ways; see discussion of "Guarantee Update Process" in Part II and Appendix A of Part III.

In order to ensure that users actually install updates to rulesets, it may be desirable for rulesets to have a fixed lifespan, such that the device will refuse to function or revert to a baseline ruleset (see below) if its ruleset has expired.

Compared to relying on verification, rules have the basic advantage that they directly prevent violations, rather than relying on some kind of oversight to identify violators and a system of punishments to deter them. This makes them potentially both more secure and privacy-preserving. On the other hand, rules are technically somewhat more challenging to implement, and whoever has the power to set the rules can also very cheaply exert great power over the uses of the devices. The complexities of these tradeoffs are discussed further in Part III.

## Binding Commitments

One middle path between verification and externally imposed rules could be binding commitments: FlexHEG devices could allow users to commit their devices to adhere to a ruleset for some fixed length of time, in such a way that even the user themselves cannot revoke this ruleset, and that the flexHEG devices can verify to other parties that the commitment has been made.

This could allow a situation that preserves much of the advantages of rulesets, without requiring the parties to subject themselves to an external rulemaking process. Different parties could agree to make particular binding commitments, if the other party makes the same commitment.[20] To prevent a situation where commitments temporarily lapse, participants in this kind of decentralized rulemaking could commit to renew their binding commitments some time before their previous binding commitment expires, thus giving other parties some warning if one party appears to be intent on reneging on their earlier commitments.

## Discretionary Enforcement Mechanisms

A ruleset would generally be the same for everyone, and thus whatever entity or group has the power to set the rules would not be able to easily exert arbitrary power over any particular actor. This is often desirable, but in some cases it may also be desirable to be able to use discretionary enforcement mechanisms that can target governance measures to a specific device or actor based on actions they have taken that are not observable on-device.

### Operating Licenses[21]

FlexHEG devices could be configured to require a cryptographic license from some authority to perform particular actions. For example, performing very large training runs or training very capable models could require a license, which would be granted if the developer in question has passed some type of audit.

This license could be time-limited, to ensure that it can effectively be revoked by refusing to issue a renewed license, e.g. if the developer has violated some regulation.[22]

Licenses might be issued to specifically allow high-risk actions. Alternatively, an unexpired license might be required even for general use of the flexHEG device, making the license requirement effectively a remote off-switch. The latter type of license requirement might be softened by always allowing some basic set of safe actions; this possibility is discussed below under the term "baseline rulesets".

In addition to requiring a license for general classes of activities, flexHEG devices could be configured to require specific actions to be approved. For example, particular types of training runs might only be allowed if that specific training run has been signed by some authority or authorities.

---

[20] This kind of meta-commitment is difficult to implement on-device, but could be navigated in practice through e.g. each party taking turns iteratively implementing a binding commitment on a subset of their devices, turning a single "round" of a coordination game into multiple rounds.
[21] We are using the term "operating license" following Aarne et al. [38]. Kulp et al. [39] referred to a similar mechanism as offline licensing.
[22] This is in fact the only reliable way to implement any kind of remote "off switch", as any positive "stop" signal could be blocked by the owner, who ultimately controls what information reaches the flexHEG device.

License mechanisms like this could be used in a top-down way by e.g. a particular country's government, or in a multilateral or distributed way where e.g. a license from any of some set of governments is sufficient. Conversely, licenses could be required to be signed by all or a majority of some set of governments, to ensure international accountability.

## Location Restrictions

The enforced equivalent of location verification are location restrictions: FlexHEG devices could be configured to go into some restricted mode if the device cannot verify that it is in an allowable location by pinging some trusted landmark.

Location restrictions are listed here under discretionary mechanisms because whatever authority controls the landmarks can in principle arbitrarily control what is considered an allowable location. If the pings from the flexHEG device to the landmark identify the device they are coming from, the landmarks could even effectively shut down a particular device by refusing to respond to their pings.

This potential for abuse could be partly mitigated in several ways. The pings from the device could be anonymized, such that the landmark would not know which device's pings to ignore[23]. Additionally, if e.g. flexHEG devices are configured to trust multiple different states' landmarks, this could help address concerns about any particular state being able to disable chips by manipulating their landmarks.

## Discretionary Mechanisms Would Ideally be Verifiable

As a kind of meta-mechanism, it could be useful in some cases for various authorities to be able to verify claims about how they have used their discretionary mechanisms, particularly what licenses they have or have not issued. For verifying that a license has been issued, the license itself is generally sufficient proof, and the recipient device can additionally verify receipt of the license. Verifying that a particular type of license has not been issued is more technically challenging, but could be useful for e.g. states verifying to other states that no operating license has been issued for a given device, and thus that device can be known to not be operational, even if the device itself cannot be inspected.

---

[23] This could potentially be undermined if the authority refuses to respond to pings that do not de-anonymize themselves.

# Baseline Rulesets

Throughout this appendix we have alluded several times to the idea that in some cases it would be useful to have a baseline ruleset that devices could revert to if e.g. its operating license is not renewed, or its ruleset has expired. Such a baseline would limit the amount of power that the authority issuing the operating license or similar would have over device owners. In this sense a baseline ruleset could be thought of as a "bill of rights" that guarantees that some forms of computation are always irrevocably allowed.

It is worth noting that it would be necessarily difficult to force users to update a baseline ruleset, meaning that any bugs or loopholes in the baseline ruleset could not reliably be fixed later. This makes them technically very challenging, as discussed under "Guarantee Update Process" in Part II.

Baseline rulesets could focus on allowing forms of computation that are either too small in scale to be dangerous[24] or are specifically known to be safe. The following subsections will discuss each of these in turn.

## Maximum Cluster Size as a Minimal Rule for Small-Scale Use Cases

If a flexHEG device can confirm that it is operating as part of a cluster that is sufficiently small that it could not be used to do anything highly dangerous, or small enough that it would be impractical to prevent malicious actors from accessing similar quantities of non-flexHEG compute, such a cluster could be assumed to be safe and thus allowed to operate even under a baseline ruleset.

This would require verifying the configuration of a cluster, including verifying that the maximum outgoing communication bandwidth of the cluster is sufficiently small, to ensure that it could not practically be used as part of a larger cluster.[25]

## Irrevocably Allowed Workloads

Specific workloads, such as running inference on a specific model with specific safeguards in place, could be cryptographically signed by authorities as "safe", such that even the baseline ruleset would allow that specific workload, even under cluster configurations on which arbitrary workloads would not be allowed. This could also be done for non-AI workloads, such as commonly run simulations.

---

[24] Even if very small-scale computations could in some cases be dangerous, bad actors intent on such computations could practically always access at least some amount of unrestricted compute, and thus imposing restrictions on small-scale computation is practically always largely futile.
[25] See discussion of technical implementation of "Encrypted Cluster Formation" in Part II. Kulp et al. [39] also discuss a very similar mechanism under the term "fixed set".

Who exactly has the power to grant these signatures, and under what conditions they would be granted, would need to be decided very deliberately. If individual states have the power to grant themselves permissions like this, it could undermine international rulesets. Additionally, irrevocable approvals of models can be risky, as the full danger and misuse potential of any given model may not be known until long after it is first deployed.

Interestingly, signing most popular models and other common workloads as "safe" could enable a situation where, by design, most chips intended to be used for inference are operating without an operating license and are limited to the baseline ruleset. In a situation like this, users might only apply for an operating license to do general purpose computing if they intend to use their chips for development rather than deployment.

## Limitations of FlexHEGs

As discussed in the main text, flexHEGs have key limitations in that:
- They can only feasibly govern the frontier of AI.
- They cannot verify many governance relevant questions about workloads, such as who the *de facto* beneficiary of a workload is.

FlexHEGs also have subtler technical limitations, which will be discussed further here. These are related to the fact that flexHEG mechanisms have limited visibility into where their inputs are coming from, and how their outputs are used, after they leave the flexHEG device or ecosystem.

If results such as trained model weights are allowed to be loaded off flexHEG devices unencrypted for use on arbitrary other compute, model weights might be further fine tuned, and in principle various results such as model weights could even combined [40], potentially resulting in systems that *de facto* exceed compute or capabilities limits that the flexHEG ecosystem had been attempting to impose. However, it appears somewhat unlikely that this could lead to capability limits being exceeded by more than what would be gained from e.g. another order of magnitude of compute capacity. Therefore this problem could be largely mitigated by being somewhat conservative when setting thresholds for what results can be moved outside the flexHEG ecosystem.

Some more intractable issues can emerge even if the release of large processed objects such as model weights to non-flexHEG devices is tightly restricted: Any system will always need to take inputs from outside the flexHEG ecosystem, and release some outputs. Even if these are restricted to be small in terms of data size, this could still allow some limits to be gamed by combining several models as "modules" of a larger system.

For example, a mixture of experts (MoE) system essentially consists of several different neural networks that are each trained to handle a particular subset of possible inputs well. Any input that the system receives is then "routed" to one or more of these experts. Because there is no significant amount of communication between the experts, guaranteeable chips would have no straightforward way of knowing that a neural network they are running or training is not in fact the entire system, but merely one of the experts in a larger system. This would allow limits on system size or capabilities to be partly circumvented. However, making the individual experts smaller will harm the performance of an MoE system and, intuitively, should limit the maximum level of intelligence the system can exhibit. This means that system size limits could still be effective if they are adjusted to appropriately account for circumvention strategies such as Mixtures of Experts.

In addition to existing MoE approaches that can pass a given input to one or more of a set of models operating in parallel, other architectures might also have more complex modular structures, such as passing the "output" of one neural network to one or more other networks, potentially many times.

It may be possible to mitigate some of these strategies somewhat by attempting to further verify where model inputs are coming from, e.g. by requiring an authenticated connection between a given individual human end user and a given neural network running on a flexHEG system. This could prevent the model developer from surreptitiously routing each human input to different modules of the system, or from feeding one neural network's output to another as if it were an input from a human user.

If necessary, FlexHEGs could also be combined with other forms of oversight to potentially address some issues like this. For example, to address worries about MoE systems, developers could be required to report basic metadata about their training runs to regulators. The regulator could then review these to look for signs of efforts to circumvent rules, such as large numbers of similar training runs being run at the same time. Requiring developers to report what each model is intended to be used for would also make it more difficult to hide these systems.